\documentstyle[psfig,11pt]{article}
\begin{document}
\vspace{1.0cm}
{\Large \bf ISOCAM PHOTOMETRY OF NARROW-LINE X-RAY GALAXIES}

\vspace{1.0cm}

J. D. Law-Green$^1$, A. Zezas$^1$, M. J. Ward$^1$, C. Boisson$^2$

\vspace{1.0cm}
$^1${\it X-ray Astronomy Group, University of Leicester, Leicester LE1
7RH, United Kingdom}\\
$^2${\it DAEC, Observatoire Paris-Meudon, 5 Place Jules Janssen, 92195
Meudon Cedex, France.}\\

\vspace{0.5cm}

\section*{{\rm ABSTRACT}}
Mid-infrared photometry of the hosts of Narrow-Line X-ray Galaxies at
6$\mu$m and 12$\mu$m has been attempted with {\it ISOCAM}. No
conclusive detections have been made. This implies that these are
quiescent objects with little or no active star-formation. Neither
X-ray binaries nor starburst-driven superwinds are consistent
explanations for the X-ray emission in these objects. We conclude that
these NLXGs are predominantly AGN-powered.\\
\addtocounter{section}{1}
\section*{{\rm INTRODUCTION}}
The spectrum of the extragalactic X-ray background (XRB) is complex
and remains poorly understood. Multiple differing populations of
sources are thought to contribute to it --- the spectrum differs
strongly from that of local AGN in the range 2--20 keV (0.4 {\it vs.}
0.7; Boldt \& Leiter 1987), suggesting that AGN may not be the
dominant contribution.

The cutoff in the X-ray source counts at $\sim 10^{-14}$ erg cm$^{-2}$
s$^{-1}$ (Barcons \& Fabian 1989) indicates that optically-selected
quasars are not sufficient to generate all of the XRB. Several authors
(e.g. Hamilton \& Helfland 1987, Griffiths \& Padovani 1990) have
suggested a faint starburst population at moderate $z$ analogous to
local {\it IRAS}-selected galaxies as a contribution to the XRB. The
X-ray emission would be generated by typically $\sim 10-100$ massive
Population I X-ray binaries (MXRB's) per galaxy.

\section*{{\rm Narrow-Line X-ray Galaxies}}

Several very deep ($S_{{\rm lim}}\sim 10^{-14}-10^{-15}$ erg cm$^{-2}$
s$^{-1}$) X-ray surveys have recently been carried out (Table 1), and
have discovered evidence for this X-ray luminous starburst population:
narrow emission-line galaxies with X-ray luminosities 10--100 times
greater than normal galaxies ({\it Narrow-Line X-ray Galaxies}, or
NLXG's).

NLXG's have narrow ($<1000$ km s$^{-1}$) emission lines of [O{\sc
ii}], [O{\sc iii}], H$\alpha$, H$\beta$, [N{\sc ii}] and [S{\sc ii}]
in common with many late-type galaxies. The line ratios (especially
[O{\sc iii}]/[O{\sc ii}]) indicate a much higher ionisation state than
in field galaxies. The [O{\sc iii}]$\lambda5007$ is often asymmetric to
the blue (Boyle et al. 1995) indicating outflow. There is little or no
reddening apparent.

Subsamples of NLXG's appear to consist of roughly similar numbers of
starburst and Seyfert galaxies, from the line ratio diagnostics
(e.g. Filippenko \& Terlevich 1992). Individual NLXGs are quite
similar in other characteristics (e.g. optical and X-ray luminosities,
redshift, optical line shapes). The population evolves strongly with
redshift, apparently as $L_{X}(z)=L_{X}(0)(1+z)^{C}$; $C=2.5\pm 1$
(Boyle et al. 1995a, Pearson et al. 1997), similar behaviour to that
for radio galaxies and radio-loud quasars, optically-selected QSOs and
X-ray selected AGN. NLXGs may contribute 15-35 per cent of the XRB,
dependent on the extrapolated shape of the luminosity function and
effects of metallicity.

The existence of X-ray luminous ($L_{X}>10^{42}$ erg s$^{-1}$)
starbursts is still controversial, however. Moran et al. (1996) find
that NLXG's in the EMSS (Stocke et al. 1991) with ambiguous
classifications have faint broad wings in their H$\alpha$ lines, and
argue that they are intermediate-type Seyferts (1.8 or 1.9) where the
broad-line emission is heavily-reddened and possibly variable. Line
diagnostics of composite galaxies may be problematic. We clearly need
additional constraints on the starburst/AGN nature of NLXG's, and so
we have conducted a deep imaging survey of several NLXG's with {\it
ISO}, in order to constrain their SED's in the mid-infrared.

\section*{{\rm OBSERVATIONS}}

Our {\it ISO} targets were chosen from three deep X-ray surveys: EMSS
(Stocke et al. 1991), CRSS (Boyle et al. 1995a), and the deep {\it
ROSAT} survey of Griffiths et al. (1996). Subsets of the brightest
X-ray targets were selected. Within the scheduling constraints of the
{\it ISO} mission, 13 targets were chosen, and 11 were actually
observed. Each target was observed with {\it ISOCAM} (C\'{e}sarsky et
al. 1996) using the LW2 (5.5--8.5$\mu$m) and LW10(8.0--15.0$\mu$m)
filters. The image scale was chosen as 3$''$ per pixel. The
observations were microscanned in a 2$\times$2 raster, offsetting
6$''$ between each point. The integration time was set as 20.16 s per
frame. A total of 20 frames were taken at each position to allow the
detectors to stabilise, then a further 20 frames for target data.  The
images were processed with the CAM Interactive Analysis (CIA) software
--- frames were corrected for dark current, glitches in both spatial
and temporal domains were searched for and removed using the
Multiresolution Median Transform algorithm, transients were removed
using the IAS algorithm, and the frames were flat-fielded, registered
and stacked to form the final mosaics.

\begin{table*}
\centering
\normalsize
\caption[]{
Selected data for {\it ISO} targets. \dag: Luminosity in {\it
Einstein} IPC band (0.3--3.5keV). Line flux ratios measured from
low-resolution (6 A) WHT spectra. References: 1: Boyle et
al. (1995a). 2: Ciliegi et al. (1995a). 3: Stocke et al. (1991). Where
$B$ magnitudes were not available, $B-V=0.3$ was assumed.\\[0.5cm]
}
\footnotesize
\begin{tabular}{lcccccccc}\hline
Source                  & Date obs. & $z$   & $m_{B}$ & $L$(0.5--2keV) & $\mathrm{\frac{I([OIII])}{I(H\beta)}}$ & 
$\mathrm{\frac{I([NII])}{I(H\alpha)}}$& ID & Ref.\\ 
                        &       &     &     & (erg s$^{-1}$) &    &   \\ \hline\hline
                      &              &       &       &                    &     &     &     &    \\
{\bf MS1414.8--1247}  & 1997 Aug 20  & 0.198 & 19.32 & $1.3\times10^{44}$\dag & 2.08 & 0.33 & HII & 3\\ 
{\bf CRSS1514.4+5627}   & 1997 Sep 1 & 0.446 & 20.69 & $3.8\times10^{43}$ & --- & --- & ?   & 1\\
{\bf CRSS1705.3+6044} & 1997 Sep 3   & 0.572 & 19.29 & $4.6\times10^{43}$ & $>2$& --- & Sy2 & 2\\
{\bf CRSS1605.6+2543} & 1997 Sep 21  & 0.278 & 18.9  & $1.7\times10^{43}$ & 0.5 & 0.4 & HII & 1\\
{\bf CRSS1605.9+2554} & 1997 Sep 21  & 0.151 & 20.55 & $4.4\times10^{42}$ & --- & --- & ?   & 1\\
{\bf CRSS0030.7+2616} & 1997 Dec 16  & 0.246 & 18.70 & $5.9\times10^{42}$ & --- & 0.78 & ?  & 1 \\
{\bf GSGP4X:48}       & 1997 Dec 3   & 0.155 & 20.37 & $3.4\times10^{42}$ & --- & --- & ?   & 4\\
{\bf GSGP4X:91}       & 1997 Dec 3   & 0.416 & 21.33 & $3.1\times10^{42}$ & --- & --- & ?   & 4\\
{\bf MS2338.9--1206}  & 1997 Dec 13  & 0.085 & 18.99 & $9.1\times10^{42}$ & 3.2 & 0.19 & Sy2/HII & 4 \\
{\bf QSF1X:36}        & 1998 Jan 15  & 0.551 & 21.07 & $3.6\times10^{43}$ & --- & --- & ?   & 4\\
{\bf MS0423.8--1247}  & 1998 Mar 19  & 0.161 & 19.76 & $8.7\times10^{43}$ & --- & 0.34 & Sy2/HII & 4\\ \hline
\end{tabular}
\end{table*}

\section*{{\rm RESULTS}}
\begin{table*}
\centering
\normalsize
\caption[]{
{\small
Results of ISOCAM photometry of Narrow-Line X-ray Galaxies. Limits are
conservative 5$\sigma$ upper limits, where $\sigma$ is the
pixel-to-pixel RMS scatter in the ISOCAM frames. $L_{{\rm fir}}$:
Far-IR luminosity, calculated using Soifer et al. (1987). $SFR:$\/
Star formation rate, extrapolated from FIR luminosity by $SFR\sim
26L_{{\rm IR,11}} M_{\odot}$ yr$^{-1}$ (Hunter \& Gallagher
 1987). $L_{X}$(superwind): X-ray emission expected from superwind
driven by starburst with observed parameters. $N_{{\rm HMXRB}}$:
number of HMXRB's expected based on IR emission from OB stars.\\[0.5cm]
}}
\footnotesize
\begin{tabular}{lcccccc}\hline
Source & $S(6\mu$m) & $S(12\mu$m) & $L_{{\rm fir}}$ & $SFR$        & $L_{X}$(superwind)      & $N_{{\rm HMXRB}}$ \\
       & ($\mu$Jy)  & ($\mu$Jy)   & (erg s$^{-1}$)  & ($M_{\odot}$ yr$^{-1}$)& (erg s$^{-1}$)&                  \\
       &            &             &                 &              &                         &                  \\ \hline\hline
{\bf CRSS1514.4+5627} & $<240$ & $<290$ & $<1.3\times10^{43}$ & $<0.9$ & $<1.1\times10^{40}$ & $<2.6$ \\
{\bf CRSS1605.6+2543} & $<190$ & $<290$ & $<5.5\times10^{42}$ & $<0.4$ & $<5.0\times10^{39}$ & $<1.1$ \\
{\bf CRSS1605.9+2554} & $<190$ & $<260$ & $<1.7\times10^{42}$ & $<0.1$ & $<1.7\times10^{39}$ & $<0.3$ \\
{\bf CRSS1705.3+6044} & $<220$ & $<310$ & $<3.7\times10^{43}$ & $<2.5$ & $<3.0\times10^{40}$ & $<7.4$ \\
{\bf MS1414.8-1247}   & $<330$ & $<770$ & $<6.3\times10^{42}$ & $<0.4$ & $<5.7\times10^{39}$ & $<1.3$ \\ \hline
\end{tabular}
\end{table*}
\setlength{\unitlength}{1cm}
\begin{center}
\begin{figure}
{\small 
\noindent{}Fig. 1: {\it (A: Left:)}\/ Comparison of NLXGs and sources from the Extended
12-Micron Sample (Rush, Malkan \& Spinoglio 1993). {\it ROSAT}\/
0.1--2.4 keV fluxes for 12 Micron sources taken from Rush et
al. (1996). Luminosities assume $H_{0}=50$ km s$^{-1}$ Mpc$^{-1}$,
$q_{0}=0.5$. {\it (B: Right:)}\/ Colour-colour plots showing ratios of $B$
magnitude to 6$\mu$m and 12$\mu$m flux for one NLXG
(CRSS1605.6+2543). Tracks represent colours of starbursts from
GISSEL96 stellar evolution library (Bruzual \& Charlot 1993).}

\begin{picture}(6.0,9.0)(0,0)
\put(0,0){\includegraphics{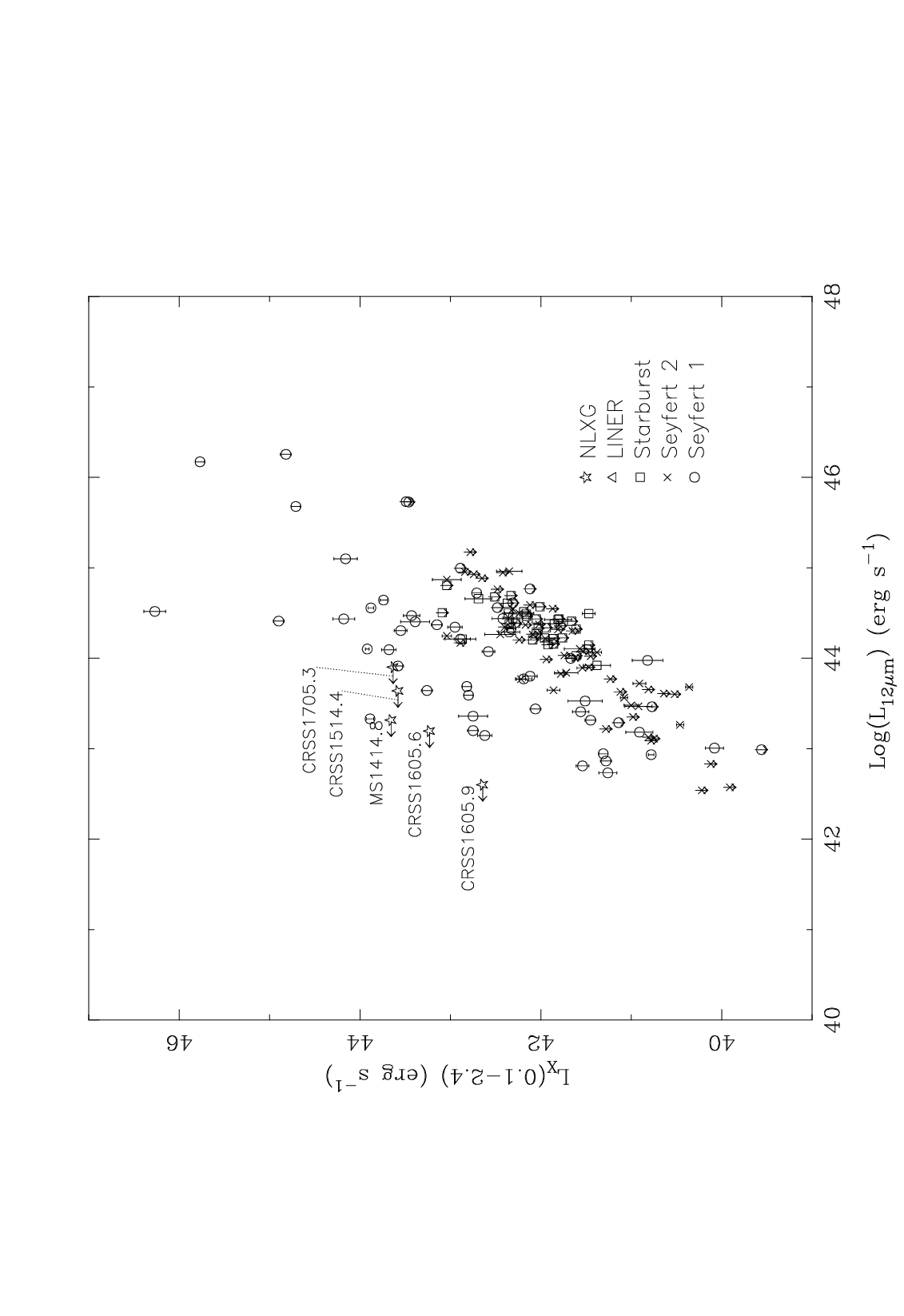}}
\put(0,0){\includegraphics{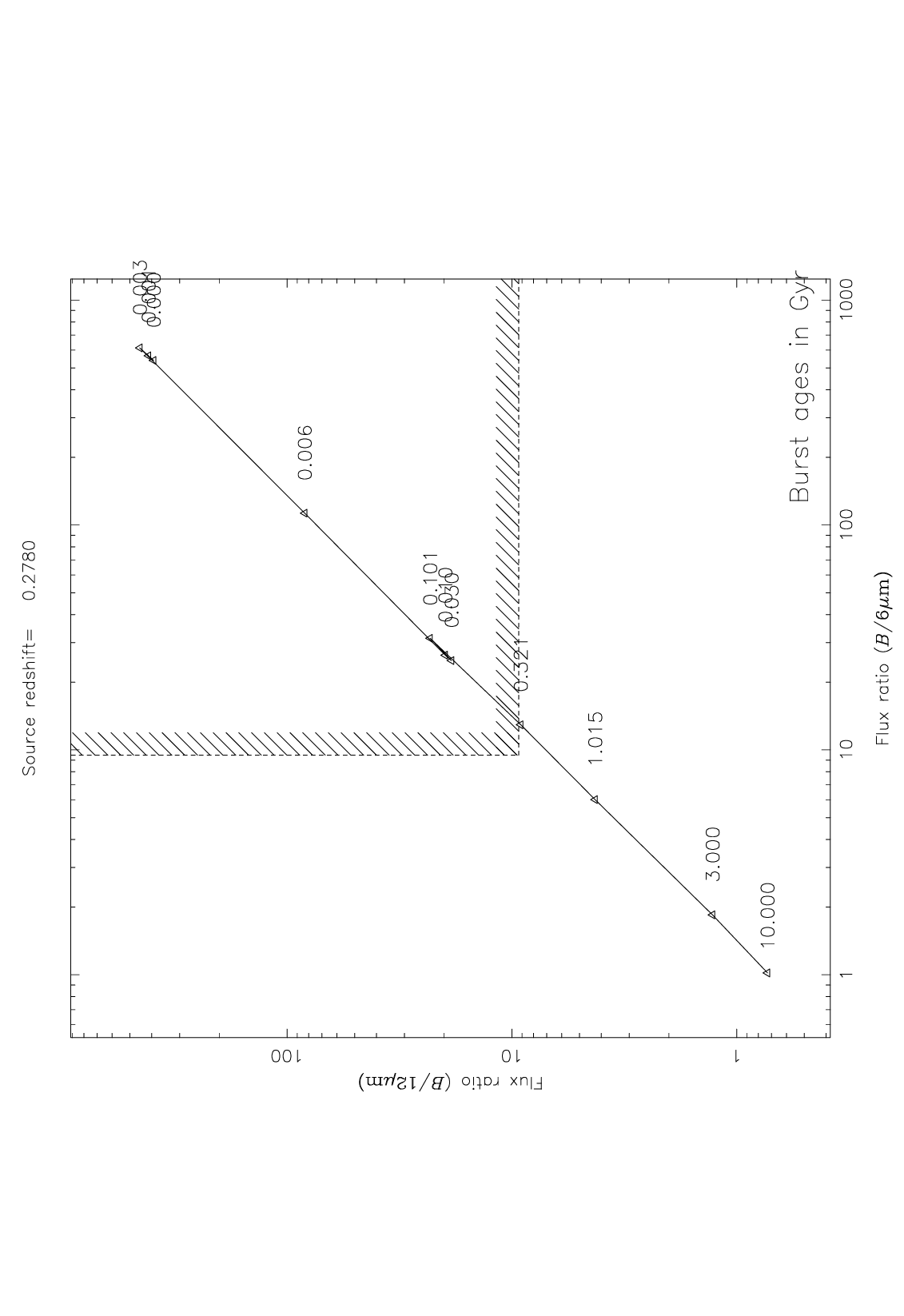}}
\end{picture}
\end{figure}
\end{center}
\normalsize
This paper discusses only those five NLXGs for which data had been
received at the time of writing. Regrettably, none of the resulting
ISOCAM frames studied so far shows any statistically reliable
detections. The upper limit on source detections is given by Table 1
as $5\sigma$, where $\sigma$ is the pixel-to-pixel RMS scatter in flux
values. In cases of possible source detection ($\geq 3\sigma$) the
temporal behaviour of the raw data was investigated, and found to be
consistent with detector glitching (sudden sharp rise followed by
exponential recovery).

The upper limits on integrated FIR luminosities were calculated using
the method of Soifer et al. (1987), using the relations between X-ray
luminosity and IR colours found by Green et al (1992). We then
followed Zezas et al. (1998) in estimating a number of the starburst
properties from the overall IR luminosity, and testing whether
they were consistent with the X-ray emission (Table 2). 

The global star formation rate $SFR$ was determined by $SFR=26L_{{\rm
IR,11}} M_{\odot}$ yr$^{-1}$ (Hunter et al. 1987). The NLXGs appear to
be quiescent galaxies, with SFR's $<2.5 M_{\odot}$ yr$^{-1}$ ({\it
cf.}  M82: 4--9 and luminous starbursts $\geq 50 M_{\odot}$
yr$^{-1}$). The IR luminosity was used to estimate the kinetic energy
deposition rate from supernovae and stellar winds, and hence the X-ray
luminosity of the gas which is associated with the starburst-driven
superwind (Heckman et al. 1996). A starburst age of 10 Myr and ambient
density of 1 cm$^{-3}$ was assumed. The ratio of these superwind
luminosities to the observed X-ray luminosities is typically very
small ($L_{{\rm X,superwind}}/L_{{\rm X,obs}} \sim 10^{-4}$), and we
conclude that superwinds are not a viable source for the soft X-ray
emission in these NLXG's.

The typical hard X-ray luminosity of high mass X-ray binaries is
$10^{37-38}$ erg s$^{-1}$. We do not have definitive hard X-ray
spectral fits on these sources, but we estimate their hard X-ray
luminosities to be $\sim 10^{42-43}$ erg s$^{-1}$. This translates
into $10^{4}$ to $10^{6}$ HMXRBs per galaxy. This can then be compared
with the number of ionising OB stars per galaxy, assuming that it is
mostly these stars which heat the dust, generating the FIR
luminosity. We estimate between 1500 and 3.5$\times 10^{4}$ per
galaxy. If $\sim 0.2$ percent of these are massive X-ray binaries
(Fabbiano et al. 1992), we obtain the (very low) numbers given in
Table 2. HMXRB's are not a viable source for the soft X-ray emission
in these NLXG's.

The presence of broad components in the H$\alpha$ lines remains
unclear from existing optical spectra, but from the above it seems
that the most likely explanation for the X-ray emission in these
sources is low-luminosity AGN activity (as in NGC 3628; Yaqoob et
al. 1995).

In principle, we should be able to constrain the properties of
possible starburst components by comparison of the observed limits on
optical-MIR colours with synthetic stellar evolution models, such as
those of Bruzual \& Charlot (1993). Figure 1(B) shows a first crude
attempt at this, showing optical/MIR tracks described by
``instantaneous'' starbursts as a function of burst age. The hatched
area shows the region within the observed optical colour limits --- in
most cases the colours are {\it not} consistent with ``old'' stellar
populations $t\geq 1$ Gyr. 

We expect to obtain the data for the remaining ISOCAM frames
shortly. Calibration of the ISO instruments is being continuously
improved in post-mission analysis. Improvements in the data-processing
may well allow the flux limits to be improved. Starburst models will
explore a range of metallicity, and AGN contributions.

\normalsize
{\it Acknowledgments:} The Infrared Space Observatory (ISO) is an ESA project
funded by ESA Member States (especially the PI countries: France,
Germany, The Netherlands and the United Kingdom), with the
participation of ISAS and NASA. CAM Interactive Analysis is a joint
development between the ESA Astrophysics Division and the ISOCAM Consortium.

\clearpage
\section*{{\rm REFERENCES}}
Barcons, X., Fabian, A. C., {\it Mon. Not. R. Astron. Soc.},
{\bf 237}, 119 (1989)\\
Boldt, E., Leiter, D., {\it Astrophys. J. Lett.}, {\bf 332}, 1 (1987)\\
Boyle B. J. et al., {\it Mon. Not. R. Astron. Soc.}, {\bf 272}, 462, (1995a)\\
Boyle B. J. et al., {\it Mon. Not. R. Astron. Soc.}, {\bf 276}, 315 (1995b)\\
Bruzual A. G., Charlot, S., {\it Astrophys. J.}, {\bf 405}, 538 (1993)\\
C\'{e}sarsky, C. J., et al., {\it Astron. \& Astrophys.}, {\bf 315},
32 (1996)\\
Ciliegi, P. et al., {\it Mon. Not. R. Astron. Soc.}, {\bf 277}, 1463 (1995)\\
Fabbiano, G. et al. {\it Astrophys. J. Suppl.}, {\bf 80}, 531 (1992)\\
Filipenko, A. V., Terlevich, R. J., {\it Astrophys. J.}, {\bf 397},
L79 (1992)\\
Green, P. J., Anderson, S. F., Ward, M. J., {\it
Mon. Not. R. Astron. Soc}, {\bf 254}, 30 (1992)\\
Griffiths, R. E., Padovani, P., {\it Astrophys. J.}, {\bf 360}, 483 (1990)\\
Griffiths, R. E. et al., {\it Mon. Not. R. Astron. Soc.}, {\bf 281},
71 (1996)\\
Hamilton, T. T., Helfland, D. J., {\it Astrophys. J.}, {\bf 318}, 93 (1987)\\
Heckman, T. M.  et al. {\it Astrophys. J.}, {\bf 457}, 616 (1986)\\
Hunter D. A., Gallagher, J. S., in {\it Star Formation in Galaxies},
ed. C. Persson (NASA CP2466), 257 (1987)\\
Moran, E. C. et al., {\it Astrophys. J. Lett.}, {\bf 433}, 65 (1994)\\
Moran, E. C., Halpern, J. P., Helfland, D. J., {\it
Astrophys. J. Suppl.}, {\bf 106}, 34 (1996)\\
Pearson, C. P. et al., {\it Mon. Not. R. Astron. Soc.}, {\bf 288},
273 (1997)\\
Soifer B. T. et al., {\it Astrophys. J.}, {\bf 320}, 238 (1987)\\
Stocke, J. T. et al, {\it Astrophys. J. Suppl.}, {\bf 76}, 813 (1991)\\
Yaqoob, T., et al., {\it Astrophys. J.}, {\bf 455}, 508 (1995)\\
Zezas, A. L., Georgantopoulos, I., Ward, M. J., in
astro-ph/9807258 (1998)\\
\end{document}